\documentclass{article}

\usepackage{arxiv}

\usepackage[utf8]{inputenc} 
\usepackage[T1]{fontenc}    
\usepackage{hyperref}       
\usepackage{url}            
\usepackage{booktabs}       
\usepackage{amsfonts}       
\usepackage{nicefrac}       
\usepackage{microtype}      
\usepackage{lipsum}         
\usepackage{amsmath}
\usepackage{cleveref}       
\usepackage{graphicx}
\usepackage{subfig}
\usepackage{multicol}
\usepackage[super,comma,sort&compress]{natbib}
\usepackage{doi}
\usepackage{xcolor}
\usepackage{lineno}
\usepackage[normalem]{ulem}
\usepackage{cancel}
\usepackage{float}
\DeclareMathOperator*{\argmax}{arg\,max}
\DeclareMathOperator*{\argmin}{arg\,min}

\title{Simplest Mechanism Builder Algorithm (SiMBA): An Automated Microkinetic Model Discovery Tool}

\author{\href{https://orcid.org/0000-0001-5273-7491}{\includegraphics[scale=0.06]{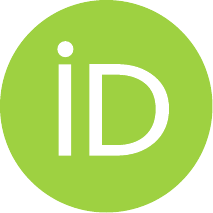}\hspace{1mm}Miguel Ángel de Carvalho Servia} \\
	Department of Chemical Engineering\\
	Imperial College London\\
    South Kensington, London, SW7 2AZ, UK\\
	\texttt{m.de-carvalho-servia21@imperial.ac.uk} \\
 \And
  \href{https://orcid.org/0000-0002-1163-0505}{\includegraphics[scale=0.06]{orcid.pdf}\hspace{1mm}King Kuok (Mimi) Hii} \\
	Department of Chemistry\\
	Imperial College London\\
    White City, London, W12 0BZ, UK\\
	\texttt{mimi.hii@imperial.ac.uk}
  \And
 \href{https://orcid.org/0000-0002-4630-1015}{\includegraphics[scale=0.06]{orcid.pdf}\hspace{1mm}Klaus Hellgardt} \\
	Department of Chemical Engineering\\
	Imperial College London\\
    South Kensington, London, SW7 2AZ, UK\\
	\texttt{k.hellgardt@imperial.ac.uk}
 \And
   \href{https://orcid.org/0000-0001-5956-4618}{\includegraphics[scale=0.06]{orcid.pdf}\hspace{1mm}Dongda Zhang} \\
    Department of Chemical Engineering\\
    The University of Manchester\\
    Manchester, M13 9PL, UK \\
    \texttt{dongda.zhang@manchester.ac.uk} 
  \And
  \href{https://orcid.org/0000-0003-0274-2852}{\includegraphics[scale=0.06]{orcid.pdf}\hspace{1mm}Ehecatl Antonio del Rio Chanona $^*$} \\
	Department of Chemical Engineering\\
	Imperial College London\\
    South Kensington, London, SW7 2AZ, UK\\
	\texttt{a.del-rio-chanona@imperial.ac.uk}
}

\renewcommand{\headeright}{}
\renewcommand{\undertitle}{}

\begin{document}
\maketitle

\begin{abstract}
Microkinetic models are key for evaluating industrial processes' efficiency and chemicals’ environmental impact. Manual construction of these models is difficult and time-consuming, prompting a shift to automated methods. This study introduces SiMBA (Simplest Mechanism Builder Algorithm), a novel approach for generating microkinetic models from kinetic data. SiMBA operates through four phases: mechanism generation, mechanism translation, parameter estimation, and model comparison. Our approach systematically proposes reaction mechanisms, using matrix representations and a parallelized backtracking algorithm to manage complexity. These mechanisms are then translated into microkinetic models represented by ordinary differential equations, and optimized to fit available data. Models are compared using information criteria to balance accuracy and complexity, iterating until convergence to an optimal model is reached. Case studies on an aldol condensation reaction, and the dehydration of fructose demonstrate SiMBA’s effectiveness in distilling complex kinetic behaviors into simple yet accurate models. While SiMBA predicts intermediates correctly for all case studies, it does not chemically identify intermediates, requiring expert input for complex systems. Despite this, SiMBA significantly enhances mechanistic exploration, offering a robust initial mechanism that accelerates the development and modeling of chemical processes. By automating microkinetic model generation from a data-first approach, SiMBA opens new avenues for future research in automated mechanism discovery.

\end{abstract}

\providecommand{\keyword}[1]{\textbf{Keywords:} #1}
\keyword{chemical reaction engineering, microkinetic model generation, automated knowledge discovery}

\pagebreak

\section{Introduction}\label{Introduction}
Microkinetic models are indispensable tools in both business and policymaking due to their ability to evaluate the efficiency of industrial processes and the environmental impact of chemicals. These models are particularly vital in sectors such as pharmaceuticals \cite{Zhou_2021,Subelzu_2020,Uzondu_2018}, petrochemicals \cite{Zhang_2023,Wang_2024,Qajar_2024}, and environmental engineering \cite{Nelson_2021,Kumari_2023,Rajamohan_2022}, where they help optimize production processes, reduce costs, and improve sustainability. For instance, in the pharmaceutical industry, microkinetic models facilitate the understanding of drug interactions and optimize synthesis pathways, accelerating drug development \cite{Gaud_2024,Juhsz_2019,Maqbool_2024}. Similarly, in environmental policy, these models provide insights into the behavior of chemical reactions, supporting the formulation of regulations and safety standards, for example the Montreal Protocol and Stockholm Convention \cite{Montreal_Convention,Lallas_2001}. By simulating the steps of chemical reactions at the molecular level, microkinetic models offer a detailed understanding of reaction mechanisms, which is essential for making informed decisions in various sectors, thus balancing economic growth with environmental protection.

Despite their importance, the manual construction of microkinetic models is a complex, time-consuming, and error-prone process \cite{Puliyanda_2022,Ratkiewicz_2005}. Traditional methods require experts to manually identify possible reaction steps and intermediates, a task that can involve analyzing hundreds of thousands of potential interactions. This meticulous process is not only slow but also susceptible to human error and often results in models that are either overly simplified or unnecessarily complex. The increasing complexity of modern chemical systems further exacerbates these challenges, highlighting the need for more efficient and reliable approaches. Consequently, there has been a significant shift towards the development of automated methods for constructing these models \cite{Liu_2021,Broadbelt_1994,Warth_2000,Vandewiele_2012,Rangarajan_2012,Ranzi_2001,Blurock_1995,Fontain_1987,Porollo_1997,Karaba_2013,Chinnick_1988}, exploiting advances in data-driven methodologies and computational resources to streamline and enhance the accuracy of the modeling process.

The general trend towards automation, or scientific machine learning, offers substantial benefits, including increased efficiency and reduced error rates, compared to traditional manual methods \cite{Peterson_2025}. Various algorithms for generating mechanisms have been developed, typically falling into two categories: combinatorial algorithms and algorithms based on reaction classes \cite{Ratkiewicz_2005}. In the former approach, the generation of the entire set of possible reactions is based solely on the congruence of the electronic configurations of reactants and products, utilizing graph theory and bond-electron matrix representations of molecules \cite{DiMaio_1992,Clymans_1984,Broadbelt_1994,Susnow_1997}. These methods can produce highly detailed and comprehensive reaction networks. The latter approach involves algorithms that, after recognizing the compounds as belonging to a certain class, generate only those reactions known to be characteristic of that class. While this method produces more compact networks, it requires prior knowledge of existing reaction classes \cite{Warth_1998,Glaude_2000,Ratkiewicz_2002,Ranzi_1995}. Combinatorial algorithms often yield overly complex mechanisms that can hinder computational efficiency and interpretability whilst making experimental validation and parameter estimation challenging (and often impossible). Conversely, algorithms based on reaction classes are limited by the availability of pre-existing reaction knowledge which may be limited within the scope of mechanism discovery for novel reactions. For a more in-depth discussion of these methodologies, reviews by \citet{Ratkiewicz_2002, VandeVijver_2015} provide valuable insights. 

In this work, we propose a new approach, the Simplest Mechanism Builder Algorithm (SiMBA), which is designed to circumvent the necessity for substantial prior knowledge required by reaction class approaches, and to avoid the proposal of overly complex mechanisms yielded by combinatorial approaches. This is done by tackling the problem of automated generation of mechanisms from a data-first perspective, ensuring that whatever mechanism is proposed, is both physically reasonable and only as complex as the data allows. SiMBA generates microkinetic models that progressively increase in complexity based on the provided data. The algorithm begins with the simplest possible mechanism, yielding the most straightforward microkinetic model. The complexity of the mechanism is then incrementally increased, thus increasing the number of parameters of the corresponding microkinetic model. This process continues as long as there is informational gain from the added parameters, which is evaluated using the Akaike Information Criterion (AIC). By balancing model simplicity and accuracy, SiMBA ensures the generation of robust and sensible microkinetic models, effectively bridging the gap between theoretical exploration and practical applicability. While alternative model discrimination measures could be employed, we chose the AIC based on prior work demonstrating its superior performance in selecting data-generating kinetic models from a set of candidates \cite{deCarvalhoServia_2023}. This minimalist approach is structurally and fundamentally different than previous methods, in that the main objective is to discover the most accurate and parsimonious mechanism given the dataset available, with as little prior information as possible. 

This research represents an advancement in the field of microkinetic modeling, offering a novel approach that overcomes many of the challenges associated with existing automated methods. By systematically generating, refining, and evaluating microkinetic models, SiMBA provides a robust framework for developing accurate and experimentally viable reaction mechanisms. The algorithm’s ability to distill complex chemical processes into simple, yet precise models has the potential to accelerate the design and optimization of chemical processes across various industries. Ultimately, SiMBA can become a useful tool for chemists and engineers, facilitating the rapid discovery and refinement of microkinetic models, thereby advancing our understanding of chemical reactions in diverse contexts.

The rest of the paper is organized as follows: in Section \ref{Methodology} our proposed method is motivated and described in detail; in Section \ref{Case Studies} we introduce the three case studies that are used to analyze the performance of SiMBA highlighting the data-generation procedure; in Section \ref{Results and Discussions} the results of the study are presented and amply discussed along with the shortcomings of the proposed methodology; and in Section \ref{Conclusions} the key findings are presented with a brief outlook on future research.

\section{Methodology}\label{Methodology}
SiMBA (Simplest Mechanism Builder Algorithm) has been tailored to develop microkinetic models using kinetic data, focusing on identifying the informationally smallest reaction mechanism that accurately describes the available data. By focusing on the balance between model accuracy and simplicity, SiMBA aims to make the process of mechanism discovery more accessible, efficient, and reliable.

SiMBA is comprised of four key phases: 

\begin{enumerate}
    \item \textbf{Mechanism generation phase}: utilizes a parallelized backtracking algorithm to generate all physically-sensible mechanisms for a given set of complexity parameters: the number of elementary steps and intermediates. This phase ensures that only feasible reaction pathways are considered, significantly reducing the computational burden;
    \item \textbf{Mechanism translation phase}: the proposed mechanisms, represented by a matrix, are converted into executable microkinetic models, specifically systems of ordinary differential equations (ODEs). This translation is crucial for transforming reaction networks into practical models that can be analyzed and simulated;
    \item \textbf{Parameter estimation phase}: the kinetic parameters of the proposed microkinetic models are estimated by minimizing the error between the model predictions and the observed kinetic data. This is achieved using the Broyden-Fletcher-Goldfarb-Shanno (BFGS) optimization algorithm;
    \item \textbf{Model comparison phase}: involves evaluating the generated models using the AIC to determine the best microkinetic model for a given iteration. This phase also decides whether further iterations and additional complexity provide enough informational gain to justify continuing the algorithm.
\end{enumerate}

By systematically progressing through these phases, SiMBA ensures the development of robust, accurate, and computationally efficient microkinetic models.

\begin{figure}[!htb]
    \centering
    \includegraphics[width=0.98\textwidth]{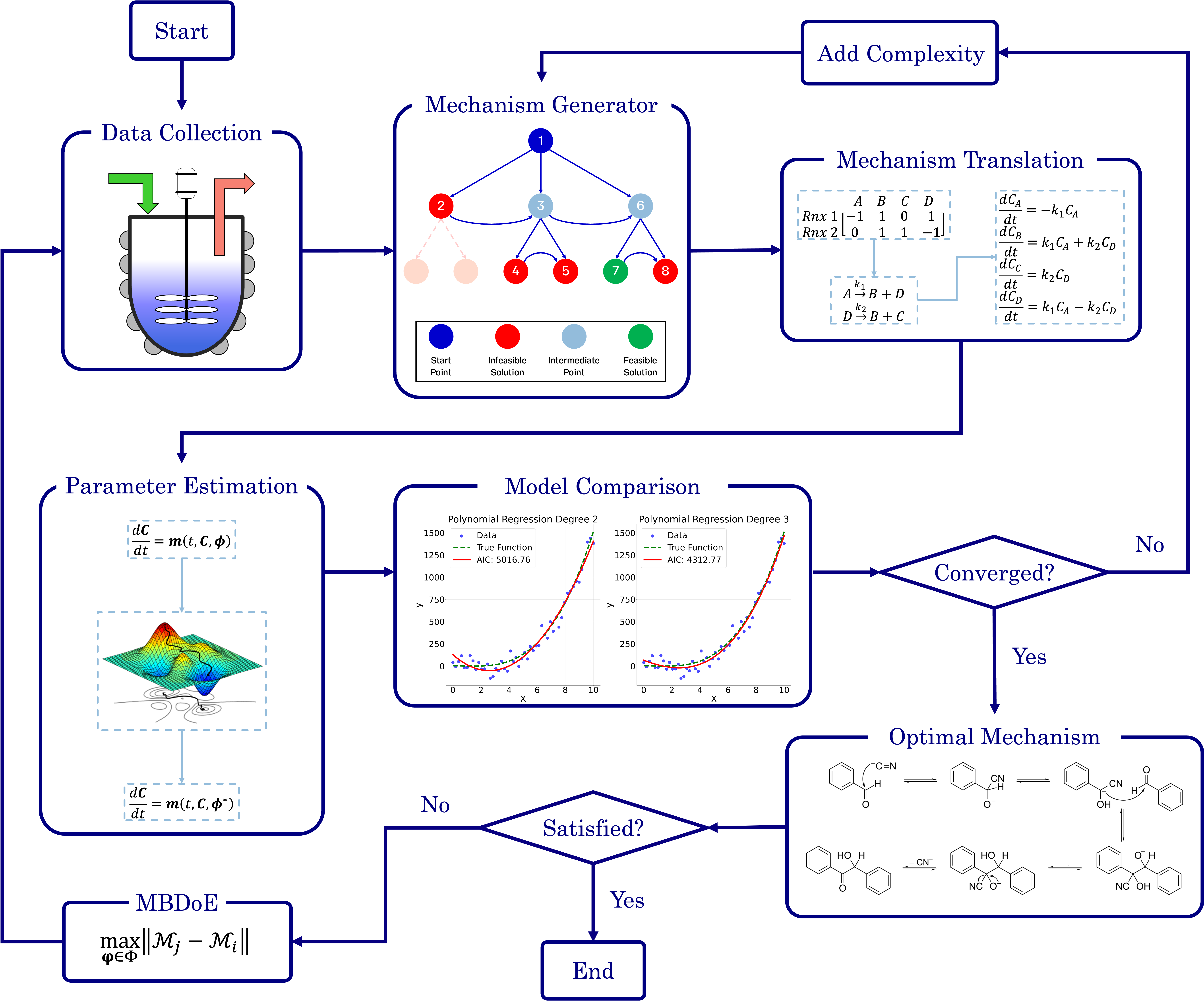}
    \caption{The workflow of the SiMBA methodology.}
    \label{Fig:nice_diagram}
\end{figure}

Our methodology also offers a closed-loop approach for refining models if the SiMBA's output is unsatisfactory, whether due to conflicts with prior knowledge (e.g., belief that the microkinetic model should involve more/less chemical species) or due to poor model fitting (e.g., the model failing to accurately capture the non-linearities in the kinetic data). In such cases, the modeler can opt to conduct an optimal experiment specifically designed to enhance model discovery -- using model-based design of experiments (MBDoE), more specifically the Hunter-Reiner criterion \cite{Hunter_1965} -- and then integrate this new data with the initial dataset. With the additional experimental data, the methodology can be re-applied, allowing for iterative refinement and re-evaluation of the output. Practically, this discriminatory experiment could also serve to validate the models proposed in earlier iterations, rather than relying solely on the AIC. The process can be repeated as many times as necessary or until the experimental budget is exhausted. Figure \ref{Fig:nice_diagram} visually represents the SiMBA workflow, highlighting the key phases of the methodology. The following subsections provide a detailed account of each of these phases.

\subsection{Mechanism Generation}\label{Generation}
The primary goal of the this initial phase is to systematically generate all physically plausible reaction mechanisms given a set of specified constraints. This phase sets the foundation for the microkinetic modeling process by exploring the potential reaction pathways that could feasibly describe the overall chemical reaction under investigation. By considering physical and chemical constraints, we ensure that only realistic and meaningful mechanisms are carried forward for further analysis.

The algorithm utilizes matrix representations to model molecular transformations, where each matrix corresponds to a potential reaction mechanism (i.e., each row accounts for an elementary step and each column accounts for a chemical species). This formalism allows the algorithm to handle complex molecular interactions in a structured manner, making it easier to apply checks and balances on the proposed mechanisms. 

SiMBA is designed to ensure that only chemically sensible and stoichiometrically balanced reactions are proposed. It does so by adhering to specific rules, which will be elaborated on later in this Section. These checks are crucial in maintaining the physical plausibility of the generated mechanisms. But before initiating the mechanism generation process, several key inputs are required:

\begin{itemize}
    \item Number of elementary reactions: this input defines the smallest possible number of reactions, or elementary steps, that could lead to a feasible microkinetic model. These elementary steps are constrained by physical principles, such as the requirement that reactions typically involve at \textbf{most} two molecules (bimolecular interactions) and usually produce a \textbf{maximum} of two product molecules (i.e., four possible elementary reactions: (i) $A \rightarrow B$, (ii) $A + B \rightarrow C$, (iii) $A \rightarrow B + C$, and (iv) $A+B \rightarrow C+D$). This consideration significantly reduces the complexity of the potential mechanisms and aligns the generated reactions with known ones.
    \item Number of chemical species: this input specifies the minimum (i.e., lower limit) number of chemical species needed to form the smallest possible mechanism. The number of species is critical because it defines the scope of the mechanism generation, ensuring that all necessary reactants, products, and intermediates are considered. This parameter also helps in maintaining the balance between complexity and feasibility in the proposed mechanisms.
    \item Stoichiometry: the stoichiometry input dictates the overall chemical reaction being analyzed. It specifies the roles of different species in the reaction, with negative numbers representing reactants, positive numbers representing products, and zeros indicating intermediates (species that do not appear in the overall reaction). This ensures that the generated mechanisms adhere to the correct chemical balance and respect the conservation of mass.
    \item Time budget: the time budget defines the amount of computational time allocated to the mechanism generator algorithm for exploring the search space and identifying physically feasible reaction mechanisms at each iteration. This constraint helps in managing computational resources effectively and ensures that the generation process is both thorough and time-efficient.
\end{itemize}

\begin{figure}[!htb]
    \centering
    \includegraphics[width=0.6\textwidth]{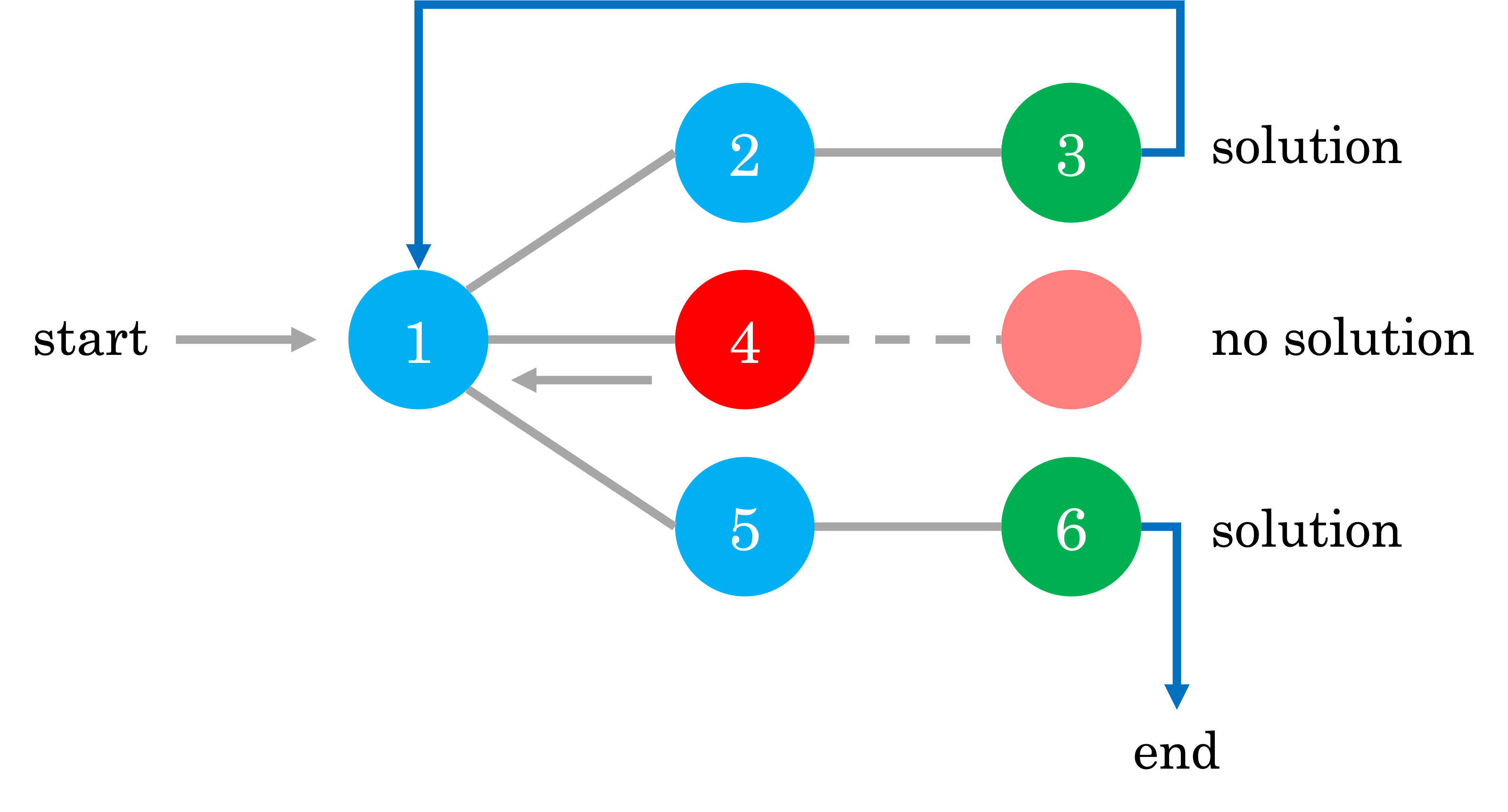}
    \caption{Example of a backtracking algorithm flowchart, where the algorithm explores potential pathways from node 1 by systematically advancing to connected nodes (2, 4, 5) while evaluating constraints. If a path fails to meet criteria (e.g., reaching a red node), the algorithm "backtracks" to the previous node, exploring alternative paths until a viable solution path is found (ending at a green node).}
    \label{Fig:backtracking_diagram}
\end{figure}

Once the inputs are defined, we can allow the backtracking algorithm to explore the mechanistic possibilities. The backtracking algorithm -- a branch-and-prune method within the field of constrained optimization -- is used to systematically explore the vast search space of possible reaction mechanisms by incrementally building potential solutions and backtracking when a solution is found to be infeasible. In the context of mechanism generation, this algorithm starts with an empty matrix representation (mechanism) and progressively starts filling the matrix with possible numbers, ensuring at each step that the proposed mechanism adheres to physical and chemical constraints, such as stoichiometry and the limits on the number of reactants and products in each step. When the algorithm encounters a dead end, where a proposed mechanism violates any of the predefined constraints, it backtracks to the previous step and tries an alternative pathway. This process allows the algorithm to efficiently prune the search space, focusing only on chemically valid and feasible mechanisms, thereby avoiding the exhaustive and brute-force approach through enumeration of all possibilities. This is of particular importance because of the combinatorial nature of the problem. For example, for a small 4x5 matrix, there are 95,367,431,640,625 possible combinations assuming that $x_{i,j} \in \{-2,-1,0,1,2\}$, where a brute-force approach would be intractable. Thus, employing smart methods for an efficient exploration of the space is paramount, even when dealing with small problems. For a more detailed discussion on backtracking, the interested reader should refer to Chapter 2 of \citet{Erickson_2019}. Figure \ref{Fig:backtracking_diagram} gives an illustrative example of how the backtracking algorithm works.

To further improve the efficiency of the exploration of the possible reaction pathways, we employ a parallelized version of the backtracking algorithm. This allows multiple `trees' or potential mechanisms to be explored simultaneously, significantly accelerating the search process. The degree of parallelization is primarily constrained by the number of available processors, making this approach highly scalable with increased computational power. Parallelizing the backtracking algorithm offers significant benefits in terms of computational efficiency and scalability. By exploring multiple potential pathways concurrently, the algorithm can cover a much larger portion of the search space within the same amount of time, making it feasible to generate comprehensive sets of candidate mechanisms even for complex reactions.

The algorithm includes several rules to ensure that the generated mechanisms are chemically plausible, to name a few:

\begin{itemize}
    \item Stoichiometric consistency: the proposed mechanisms must adhere to the stoichiometry defined by the input, ensuring that the overall reaction remains balanced.
    \item Elementary step constraints: each elementary step is restricted to having at most two reactants and two products, with at least one of each, reflecting the typical nature of an elementary step and ensuring that there is no redundant elementary steps (i.e., a row full of zeros).
    \item Intermediate formation: intermediates must be generated in the reaction network before they are consumed, maintaining a logical and sequential flow of the reaction mechanism.
\end{itemize}

These rules allow for SiMBA to filter out unfeasible or non-physical mechanisms, ensuring that the outputs are not only mathematically valid but also chemically meaningful. Thus, at the end of this phase, a comprehensive set of candidate reaction mechanisms is generated, each represented by a matrix and each having the same level of complexity (i.e., same number of elementary steps and chemical species). These matrices serve as the basis for further analysis in subsequent phases of the SiMBA methodology. This phase lays the groundwork for the rest of the SiMBA methodology, providing a robust and physically plausible set of reaction mechanisms that will be refined, validated and compared in the subsequent phases.

\subsection{Mechanism Translation}\label{Translation}
The purpose of this step of the SiMBA algorithm is to convert matrix representations of mechanisms into executable models that can be used for parameter estimation, simulation, and analysis. The translation process consists of two main steps: first, converting the matrix representation into reaction strings, and second, converting these reaction strings into systems of ODEs which are executable in Python.

In the matrix representation, each row corresponds to an elementary reaction, while each column represents a chemical species. The elements within the matrix indicate stoichiometric coefficients: negative for reactants, positive for products, and zero for species not involved in the elementary step.

In the first step, each reaction string is generated by identifying the reactants, products and the stoichiometric coefficients for every row of a given matrix. For example, the below matrix would be converted into the following reaction strings, where $A$ and $B$ are reactants, $C$ is a product and $D$ is an intermediate:

\begin{equation}
\begin{bmatrix}
-1 & -1 & 0 & 1 \\
0 & -1 & 1 & -1 
\end{bmatrix}
\left\}
\begin{aligned}
& A + B \xrightarrow{k_1} D \\
& B + D \xrightarrow{k_2} C
\end{aligned}
\right.
\end{equation}

Using mass-action kinetics, the reaction strings are then converted into ODEs. The rate of a reaction is proportional to the product of the concentrations of the reactants. For example, for the reaction string $A + B \xrightarrow{k_1} D $ the rate equation is expressed as $r = k_1C_AC_B$, where $k_1$ is the rate constant. Our code automates the translation of these reaction strings into ODEs by systematically identifying unique species, constructing rate equations, and assembling the differential equations into a comprehensive kinetic model. For instance, the system of ODEs for the above matrix representation is:

\begin{equation}\label{Eq:mech_exam}
\begin{matrix}
A + B \xrightarrow{k_1} D \\
B + D \xrightarrow{k_2} C
\end{matrix}
\left\}
\begin{aligned}
&\frac{dC_A}{dt} = -k_1C_AC_B \\
&\frac{dC_B}{dt} = -k_1C_AC_B - k_2C_BC_D \\
&\frac{dC_C}{dt} = k_2C_BC_D \\
&\frac{dC_D}{dt} = k_1C_AC_B - k_2C_BC_D
\end{aligned}
\right.
\end{equation}

The translation process is automated using a Python script, primarily leveraging the regex library \cite{VanRossum_2020}. Regex is employed to parse and extract key components from reaction strings, such as chemical species and reaction operators. Specifically, regex identifies species by matching patterns of letters and helps distinguish between different parts of the reaction strings, including reactants, products, and the reaction arrow ($\rightarrow$). This parsing ensures accurate separation and interpretation of reactants and products, facilitating the automated construction of corresponding rate equations in the ODE system.

Given the potential to generate a vast number of models in any iteration of SiMBA, automating the translation from matrix notation to executable Python functions is crucial. Manual conversion would be impractical, if not impossible, due to the sheer volume of candidate mechanisms. Therefore, this automated approach not only improves efficiency but also ensures that the subsequent phases of the SiMBA algorithm can proceed smoothly. Much of the code for the translation of reaction strings to systems of ODEs has been adapted from the work of \citet{Jiscoot_2023}.

\subsection{Parameter Estimation}\label{Estimation}
The objective of this aspect of SiMBA is to determine the kinetic parameters that best fit the generated models to the available data. This step is crucial for ensuring that the proposed reaction mechanisms are optimized so that they reflect, as accurately as possible, the observed dynamics of the chemical system. This will enable the algorithm to compare different models fairly in the next phase. The parameter estimation step is a standard procedure in model building frameworks. 

To solve the parameter estimation problem, we use simulated concentration-time profiles as the dataset. These profiles provide time-series data of species' concentrations, which are critical for fitting the kinetic models.

The parameter estimation problem is defined in Eq. \eqref{Eq:PE}, where $\hat{y}_m^{(i)}$ denote the prediction of a value coming from a proposed model $m$ at a given time $t^{(i)}$ (i.e., $\hat{y}_m^{(i)} = m(t^{(i)}\mid\theta_m)$), and $y^{(i)}$ represents the target value at a given time $t^{(i)}$ (i.e., in-silico data, in this study). Furthermore, $SSE$ represents the sum of squared errors and $n_t$ is defined as the sampling times, which are set within the fixed time interval, $t^{(i)} \in \Delta t$ where $\Delta t = [t_0, t_f]$.

\begin{equation} \label{Eq:PE}
    \theta_m^*=\argmin_{\theta} \sum_{i=1}^{n_t}{SSE\left(\hat{y}_{m}^{(i)}, y^{(i)}\right)}.
\end{equation}

The Limited-memory Broyden-Fletcher-Goldfarb-Shanno (L-BFGS) algorithm is employed for solving the parameter estimation problem \cite{Liu_1989}. L-BFGS is well-suited for handling this problem due to its performance in tasks pertaining to parameter estimation and optimization \cite{Malouf_2002, Liu_1989}.

To ensure a thorough exploration of the parameter space, we may use expert-informed or random initial guesses for the parameters, with bounds set within the range $[0,10]$ to maintain physically meaningful values in the chosen case studies (this can be changed on as-needed basis). The stopping criteria for the optimization are left to the default options in the Scipy package \cite{2020SciPy-NMeth}, and a multi-start approach is employed, where multiple runs are initiated with different starting points, and the best solution is retained.


\subsection{Model Comparison}\label{Comparison}
SiMBA uses an information criterion approach for model selection rather than a data-splitting approach, enabling the entire dataset to be used for model construction while still providing a robust and reliable method for testing the proposed models. This is especially advantageous in low-data scenarios, as it ensures that we make full use of the available information when identifying suitable microkinetic models. 

We use the Akaike Information Criterion (AIC) because in previous work we compared various information criteria to determine if any offered superior performance. We found that AIC consistently outperformed other criteria in the context of kinetic discovery; further details of these studies can be found in \citet{deCarvalhoServia_2023}.

Given a model $m$ with parameters $\theta_m$ of dimension $d_m$, the AIC is defined as:

\begin{equation} \label{eq:AIC}
    \text{AIC}_m = 2NLL(\mathbf{\theta}_m\mid\mathcal{D}) + 2d_m, 
\end{equation}

where $NLL$ represents specifically the negative log-likelihood \cite{Akaike_1974}. Given two competing models, $m_1$ and $m_2$, the preferred model would be the one with the lowest AIC value calculated by Eq. \eqref{eq:AIC}.

If iteration $n+1$ in the SiMBA algorithm results in an improvement of the AIC value compared to iteration $n$, SiMBA will continue running, further refining the model output by considering more complicated mechanisms. This approach ensures that the algorithm is consistently moving towards a model that better balances complexity with goodness of fit. However, if the best model in iteration $n+1$ displays a worsening in the AIC value compared to the best model in iteration $n$, indicating that the model has become less optimal, SiMBA will terminate the process. In this scenario, the algorithm concludes that additional iterations are implausible to produce a superior model, and it will return the best solution found during iteration $n$, which is considered the most accurate and parsimonious model according to the AIC evaluation.

\subsection{Model-Based Design of Experiments}\label{MBDoE}
If the dataset used for mechanism discovery is insufficient to yield an adequate model, and provided that the experimental budget has not been exhausted, we can use the insights from the optimized models to design a more informative experiment. Specifically, we can identify operating conditions that maximize the difference between the state predictions $\mathbf{X}$ of the two best proposed models, $\nu$ and $\mu$ based on the existing dataset. The rationale behind using the two best proposed models is discussed in \citet{deCarvalhoServia_2024}. The MBDoE approach adopted in this work was developed by \citet{Hunter_1965}:

\begin{align}\label{Eq: MBDoE}
    \mathbf{x}_{k+1} &= \argmax_{\mathbf{x}\in\mathcal{X}} \; \sum_{i=1}^T \sum_{j=1}^d (\mathbf{X}^{\nu}_{i,j} - \mathbf{X}^{\mu}_{i,j})^2 \\
    \mathbf{X}^{\nu} &= \int_{t_0}^{t_f} f^{\nu}(\mathbf{x},t,\theta^{\nu}) \; \text{d}t \\
    \mathbf{X}^{\mu} &= \int_{t_0}^{t_f} f^{\mu}(\mathbf{x},t,\theta^{\mu}) \; \text{d}t 
\end{align}


In this equation, $\textbf{x}$ represents the operating conditions within a set $\mathcal{X}$. Using the identified initial conditions, a new experiment can be conducted to generate additional data points, which are then added to the original dataset. With this updated dataset SiMBA can be executed again, thereby closing the loop between informative experimental design and optimal model discovery.

\section{Catalytic Kinetic Case Studies}\label{Case Studies}
The purpose of the case studies presented in this work is to serve as proof-of-concept validations for the newly developed methodology, SiMBA. Before deploying this data-driven method in experimental environments or attempting to propose and discover new reaction mechanisms, it is important to ensure that the methodology is both sound and capable of delivering reliable results. To achieve this, we selected case studies where experimentalists have already proposed mechanisms or rate models, allowing us to generate in-silico data through computational simulations and subsequently test SiMBA's ability to accurately rediscover these mechanisms from the generated datasets.

The selection of the case studies was made to demonstrate SiMBA’s effectiveness across a range of scenarios. The case studies include a hypothetical reaction, an aldol condensation between benzaldehyde and acetophenone \cite{Nielsen_2011}, and the dehydration of fructose to 5-hydroxymethylfurfural (HMF) \cite{vanPutten_2013,Chen_2024}. These studies were chosen to showcase SiMBA's ability to distill complex kinetic behaviors into simple, accurate models. 

The hypothetical reaction serves as an initial proof-of-concept, illustrating whether SiMBA can generate microkinetic models purely from fundamental principles without relying on prior knowledge. This first case study also demonstrates SiMBA’s versatility in handling both first-order and second-order elementary steps. Next, the aldol condensation -- a classic reaction with a well-understood mechanism -- tests SiMBA’s ability to reconstruct mechanistic pathways solely from dynamic data on main reactants and products. By successfully modeling this reaction, SiMBA shows that it can go beyond hypothetical examples to accurately capture established mechanistic pathways. Finally, the dehydration of fructose to HMF introduces a different challenge: rather than relying on a microkinetic simulation of stoichiometric reactants and products, the in-silico dataset comes from a rate model that has been experimentally validated. In this scenario, SiMBA is challenged to discover a plausible kinetic mechanism, aligning with established literature and demonstrating its capacity to derive robust reaction models from realistic data sources.



\subsection{The Hypothetical Reaction}\label{Hypothetical}
The hypothetical reaction is one that involves five different chemical species -- only three of which are observed -- interacting in four different elementary steps. The overall stoichiometry of the hypothetical reaction can be represented by Eq. \eqref{Eq:stoich_hypo} while Eq. \eqref{Eq:mech_hypo} provides a description of the mechanism of the reaction as well as the system of ordinary differential equations (ODEs) underpinning the dynamics of the reaction (and directly derived from the proposed mechanism). The kinetic parameters (rate constants) were defined as: $k_1=0.1$ M$^{-1}$ h$^{-1}$, $k_2=0.2$ h$^{-1}$, $k_3=0.13$ h$^{-1}$ and $k_4=0.25$ M$^{-1}$ h$^{-1}$.

\begin{gather}\label{Eq:stoich_hypo}
    4A \rightarrow B + C
\end{gather}

\begin{equation}\label{Eq:mech_hypo}
\begin{matrix}
2A \xrightarrow{k_1} B \\
A \xrightarrow{k_2} D \\
D \xrightarrow{k_3} E \\
A + E \xrightarrow{k_4} C
\end{matrix}
\left\}
\begin{aligned}
&\frac{dC_A}{dt} = -k_1C_A^2 - k_2C_A - k_4C_AC_E \\
&\frac{dC_B}{dt} = k_1C_A^2 \\
&\frac{dC_C}{dt} = k_4C_AC_E \\
&\frac{dC_D}{dt} = k_2C_A + k_3C_D \\
&\frac{dC_E}{dt} = k_3C_D - k_4C_AC_E
\end{aligned}
\right.
\end{equation}

Starting from the ODE system in Eq. \ref{Eq:mech_hypo}, an in-silico dataset is generated wherein $\Delta t = [0, 10]$ h and $n_t = 30$. This dataset is composed of five different experiments, each ran at different initial conditions (in molar units: $(C_{A}(t=0), C_{B}(t=0), C_{C}(t=0), C_{D}(t=0), C_{E}(t=0)) \in \{(10, 0, 2, 0, 0), (10, 2, 0, 0, 0), (10, 2, 2, 0, 0), (5 , 0, 0, 0, 0), (10, 0, 0, 0, 0)\}$); these experiments were randomly picked from a $2^k$ factorial design \cite{Mee_2009}. 

For all experiments, the system is assumed to be both isochoric and isothermal, and Gaussian noise is added to the in-silico measurements to simulate a chemical experiment. The added noise had zero mean and a standard deviation of 0.15 for $A$, $B$ and $C$. To further approximate a realistic system, we assume that we cannot measure the intermediates $D$ and $E$. The generated dataset for the one of the experiments are presented in Fig. \ref{Fig:exp_case_studies} (a). The dataset, providing 150 datapoints, has a realistic size for kinetic studies \cite{Schrecker_2023, Waldron_2020, Taylor_2021}, especially considering recent advancements in high-throughput setups. 

\subsection{The Aldol Condensation Reaction}\label{Condensation}
\begin{figure}[H]
    \centering
    \includegraphics[width=0.8\textwidth]{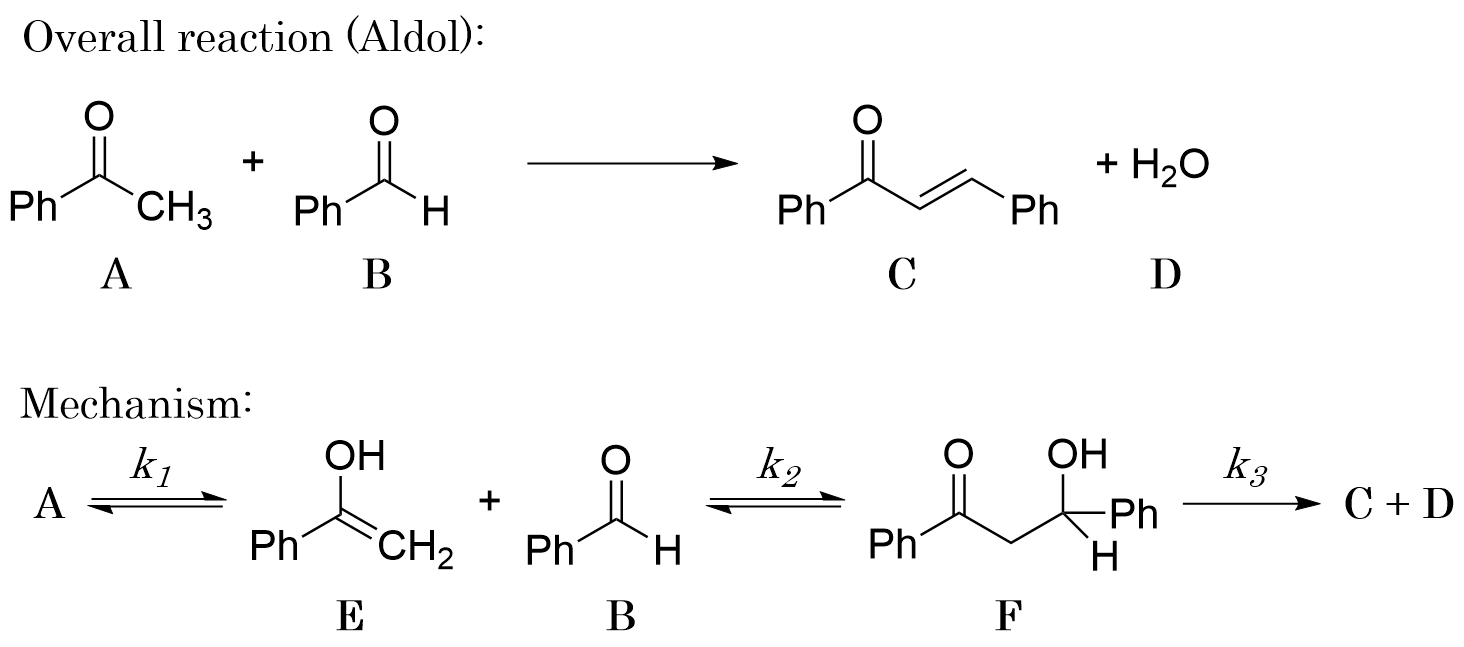}
    \caption{Schematic representation of the aldol condensation reaction between acetophenone ($A$) and benzaldehyde ($B$) to form the chalcone product ($C$) and water ($D$). The mechanism proceeds in three main steps: (i) enolization of $A$ to give the enolate/enol intermediate ($E$), (ii) nucleophilic addition of $E$ to $B$ to form the $\beta$-hydroxy adduct ($F$), and (iii) dehydration to yield the final conjugated enone ($C$). Rate constants $k_1$, $k_2$, and $k_3$ are associated with each step. Phenyl groups are represented by “Ph.”}
    \label{Fig:aldol_condensation_chem}
\end{figure}

An aldol condensation is a type of condensation reaction in organic chemistry between a ketone and an aldehyde to form a carbon-carbon double bond in the enone product, eliminating a molecule of water. The mechanism of the reaction is initiated with the formation of an enol or enolate intermediate from a ketone. This nucleophilic intermediate attacks the carbonyl group of the aldehyde, to form a $\beta$-hydroxyaldehyde or $\beta$-hydroxyketone, which in turn dehydrates to produce a conjugated enone. The aldol condensation therefore involves six different chemical species -- only four of which are observed -- interacting in three different elementary steps. Fig. \ref{Fig:aldol_condensation_chem} represents the overall reaction as well as the detailed mechanism of the aldol condensation reaction between acetophenone and benzaldehyde.

Eq. \eqref{Eq:mech_conden} provides a simplified description of the mechanism of the reaction as well as the ODE system underpinning the dynamics of the reaction (which was directly derived from the proposed mechanism) \cite{Nielsen_2011}. It is worth noting that we assume that the elementary steps are all irreversible. The kinetic parameters (rate constants) were defined as: $k_1=0.759$ h$^{-1}$, $k_2=0.293$ M$^{-1}$ h$^{-1}$ and $k_3=0.681$ h$^{-1}$. 

\begin{equation}\label{Eq:mech_conden}
\begin{matrix}
A \xrightarrow{k_1} E \\
E + B \xrightarrow{k_2} F \\
F \xrightarrow{k_3} C + D 
\end{matrix}
\left\}
\begin{aligned}
&\frac{dC_A}{dt} = -k_1C_A \\
&\frac{dC_B}{dt} = -k_2C_EC_B \\
&\frac{dC_C}{dt} = k_3C_F \\
&\frac{dC_D}{dt} = k_3C_F \\
&\frac{dC_E}{dt} = k_1C_A - k_2C_EC_B \\
&\frac{dC_F}{dt} = k_2C_EC_B - k_3C_F 
\end{aligned}
\right.
\end{equation}

In Eq. \eqref{Eq:mech_conden}, $A$, $B$, $C$, $D$, $E$ and $F$ correspond to acetophenone, benzaldehyde, chalcone, water, $\alpha$-phenylvinyl enolate and 4-hydroxy-1,3-diphenylbutan-1-one, respectively. 

Starting from the ODE system in Eq. \ref{Eq:mech_conden}, an in-silico dataset is generated wherein $\Delta t = [0, 10]$ h and $n_t = 30$. This dataset is composed of five different experiments, each ran at different initial conditions (in molar units: $(C_{A}(t=0), C_{B}(t=0), C_{C}(t=0), C_{D}(t=0), C_{E}(t=0), C_{F}(t=0)) \in \{(5 , 10, 0, 0, 0, 0), (5 , 5 , 2, 0, 0, 0), (5 , 10, 0, 2, 0, 0), (10, 10, 0, 2, 0, 0), (10, 10, 2, 2, 0, 0)\}$); these experiments were randomly picked from a $2^k$ factorial design \cite{Mee_2009}. 

For all experiments, the system is assumed to be both isochoric and isothermal, and Gaussian noise is added to the in-silico measurements to simulate a chemical experiment. The added noise had zero mean and a standard deviation of 0.15 for $A$, $B$, $C$ and $D$. To further approximate a realistic system, we assume that we cannot measure the intermediates $E$ and $F$. The generated dataset for one of the experiments are presented in Fig. \ref{Fig:exp_case_studies} (b). 

\subsection{The Dehydration of Fructose}\label{Dehydration}
The dehydration of fructose refers to the process of removing water molecules from fructose to produce 5-hydroxymethylfurfural (HMF), a valuable platform chemical. This reaction is important because HMF can be further converted into various high-value chemicals and biofuels, making it a crucial step in the conversion of biomass into renewable energy and materials. The overall stoichiometry of the dehydration reaction can be represented by Eq. \eqref{Eq:stoich_dehydr} whilst Eq. \eqref{Eq:mech_dehydr} shows the rate model extracted from the literature \cite{vanPutten_2013}, which is derived from experimental data and governs the reaction dynamics. A brief note on assumptions: the energy balance is not included in this study, instead treating the reaction as if it proceeds isothermally at 137 $^{\circ}$C. This choice reflects the experimental setup -- heating a 0.5 mL reaction mixture in sealed glass ampoules -- where the small volume and thin walls likely minimize heat-up time and heat transfer limitations. Under these conditions, the parameters used were: $C_{acid} = 3.3 \times 10^{-2}$ M of sulfuric acid, $k_{ref} = 0.9$ M$^{-1}$ min$^{-1}$, $E_a = 124$ J mol$^{-1}$, $R = 8.314$ J K$^{-1}$ mol$^{-1}$ and $T = 410.15$ K (directly taken from \citet{vanPutten_2013}). 

\begin{gather}\label{Eq:stoich_dehydr}
    A \rightarrow 3B + C
\end{gather}

\begin{subequations}\label{Eq:mech_dehydr}
\begin{align}
    r &= kC_AC_{acid} \\
    k &= k_{ref}\exp\left(-\frac{E_a}{RT}\right)
\end{align}
\end{subequations}


In Eq. \eqref{Eq:stoich_dehydr}, $A$, $B$ and $C$ correspond to fructose, water and HMF respectively. Starting from the rate model in Eq. \ref{Eq:mech_dehydr}, we can derive an ODE system ($r=\frac{1}{\nu_i}\frac{dC_i}{dt}$) and generate an in-silico dataset wherein $\Delta t = [0, 90]$ min and $n_t = 30$. This dataset is composed of five different experiments, each ran at different initial conditions (in molar units: $(C_{A}(t=0), C_{B}(t=0), C_{C}(t=0)) \in \{(4, 0, 0), (6, 2, 1), (4, 2, 0), (4, 0, 1), (6, 2, 0)\}$); these experiments were randomly picked from a $2^k$ factorial design \cite{Mee_2009}. 

\begin{figure}[htb!]
    \begin{minipage}{0.5\textwidth}
        \centering
        \subfloat[]{\includegraphics[width=\textwidth]{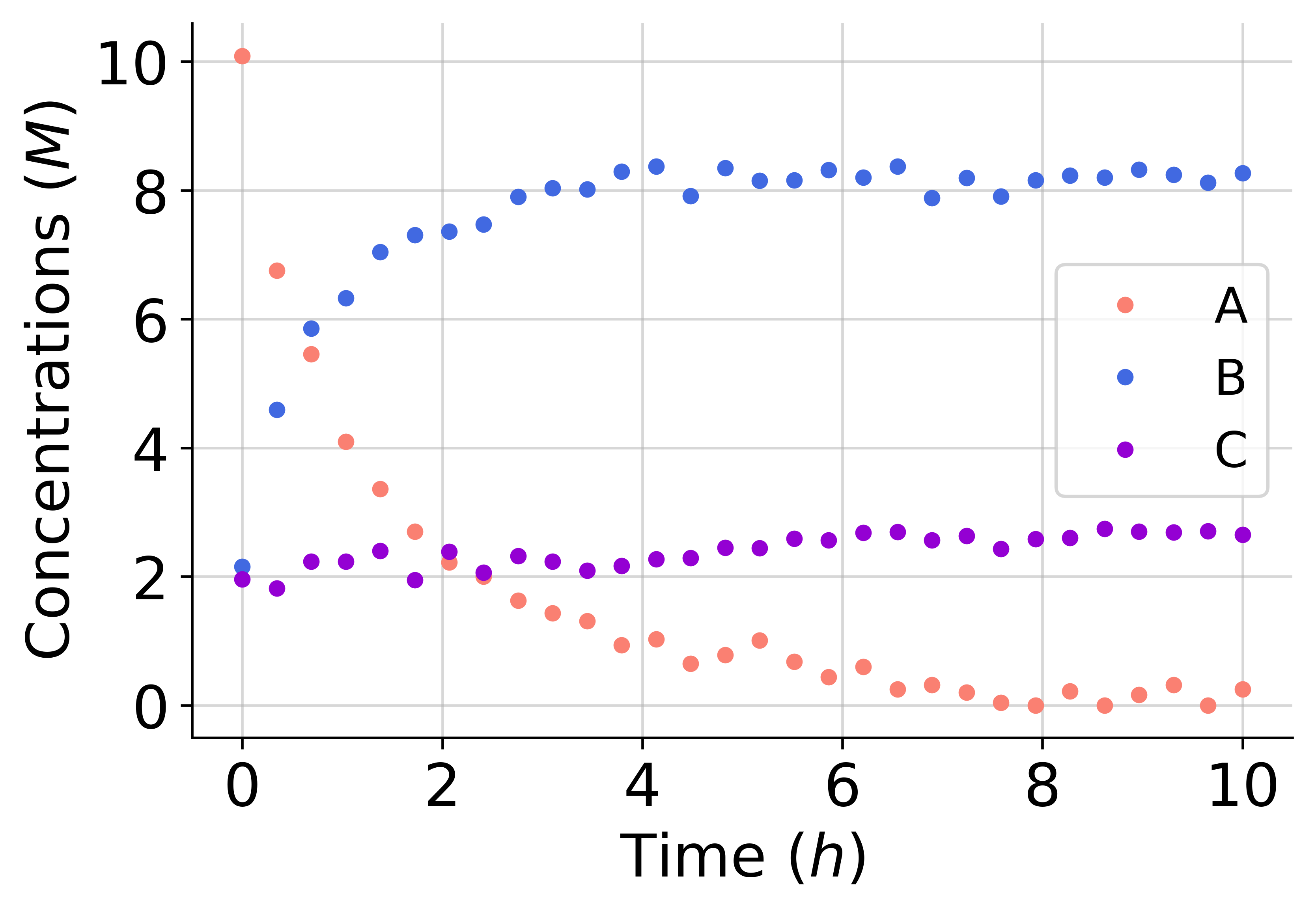}}
    \end{minipage}%
    \begin{minipage}{0.5\textwidth}
        \centering
        \subfloat[]{\includegraphics[width=\textwidth]{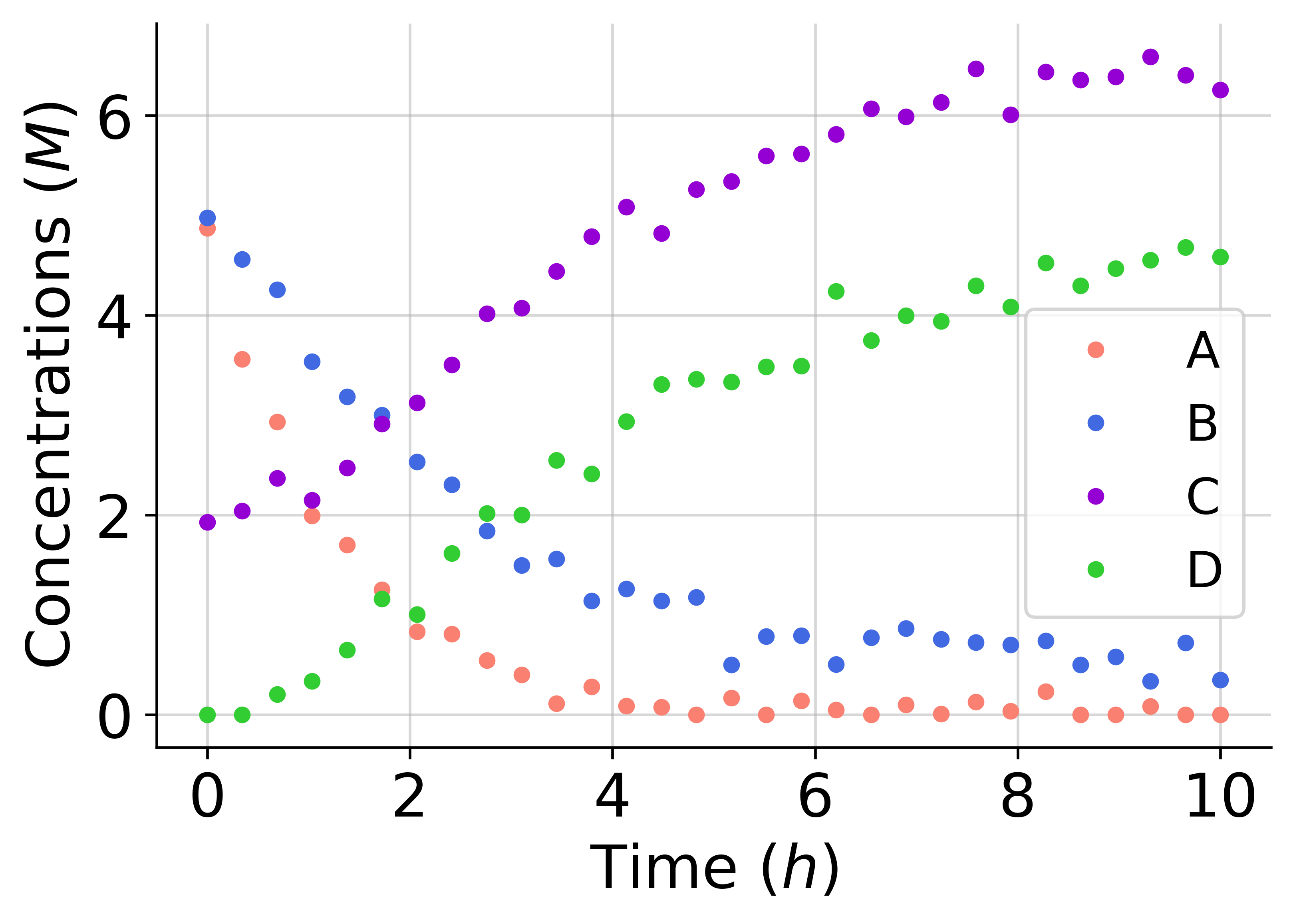}}
    \end{minipage}\par
    \centering
    \subfloat[]{\includegraphics[width=0.5\textwidth]{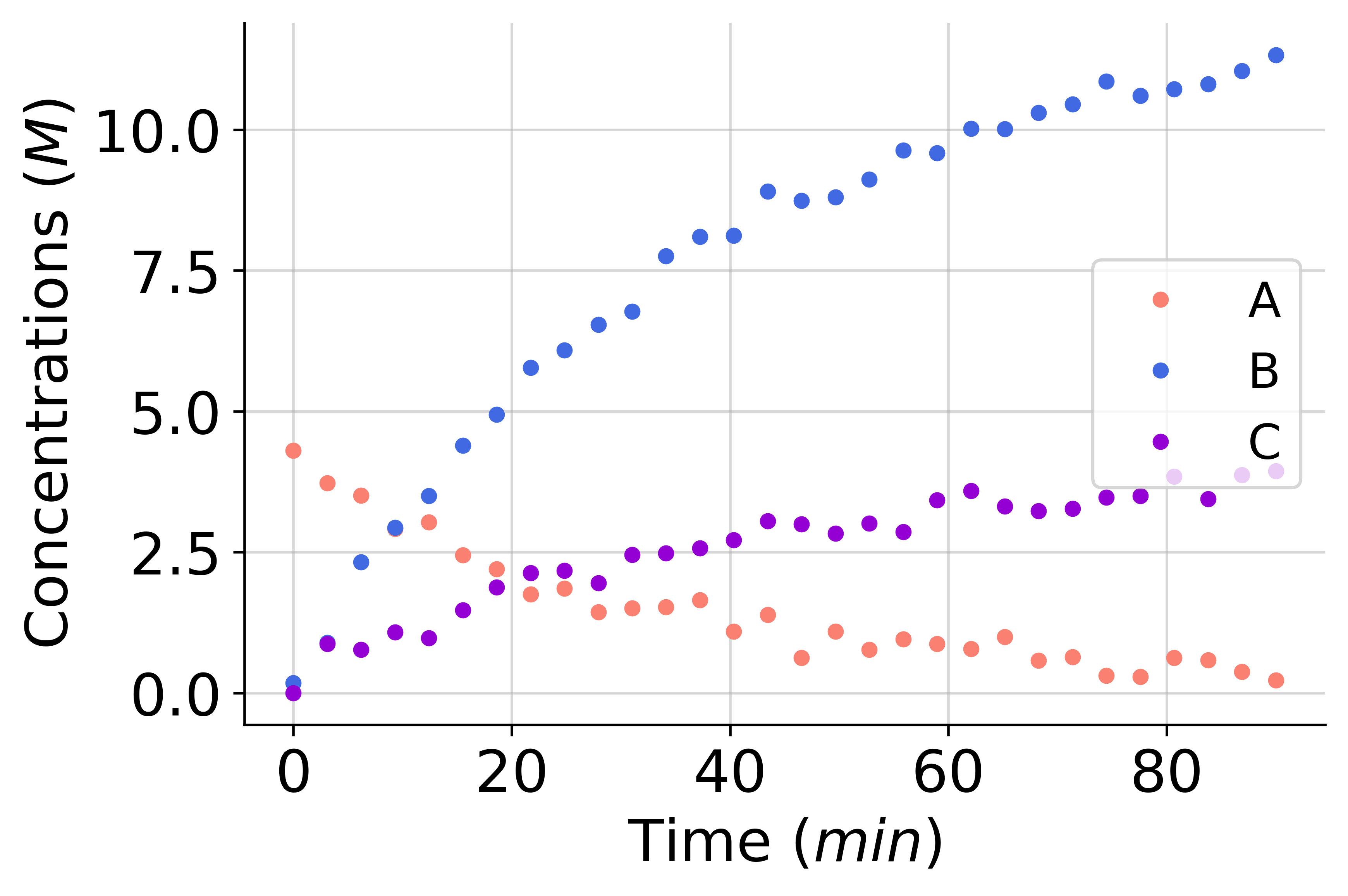}}
\caption{(a) The in-silico data of one of the computational experiments for the hypothetical reaction. (b) The generated data of one of the computational experiments for the aldol condensation reaction. (c) The generated data of one of the computational experiments for the the dehydration of fructose to HMF.}
\label{Fig:exp_case_studies}
\end{figure}

For all experiments, the system is assumed to be both isochoric and isothermal, and Gaussian noise is added to the in-silico measurements to simulate a chemical experiment. The added noise had zero mean and a standard deviation of 0.2 for $A$, $B$ and $C$. In this example, resembling a real system, we do not have any measurement on possible intermediates. The generated dataset for the one of the experiments are presented in Fig. \ref{Fig:exp_case_studies} (c). 

\section{Results and Discussions}\label{Results and Discussions}
\subsection{The Hypothetical Reaction}

The application of SiMBA to the hypothetical reaction case study successfully demonstrated its ability to recover the underlying microkinetic model with limited data and only access to the species present in the overall reaction (i.e., without direct data on intermediates). This provides initial validation of the algorithm’s capacity to propose accurate and physically sensible models under constrained conditions, highlighting its potential for broader application in more complex chemical systems, which will be shown in subsequent subsections.

In this case, given the stoichiometry of the hypothetical reaction, shown in Eq. \eqref{Eq:stoich_hypo}, the simplest possible mechanism involves two elementary steps. This is due to the fact that termolecular, and higher order interactions, are relatively rare -- for the purpose of SiMBA, we consider them as impossible -- as the simultaneous collision of three or more molecules in the correct orientation is a very unlikely occurrence. Thus, based on that constraint, we would need at least two elementary steps to react four moles of A. As such, in the first step, two moles of species A react to produce one mole of B, while in the second step, two moles of A react to produce one mole of C. Notably, the order in which B and C are produced is interchangeable, without affecting the model’s performance. These two configurations represent the only physically feasible mechanisms that could be formed in the first iteration of SiMBA, based on a 2x3 matrix (representing two elementary steps and three species).

Upon identifying all possible permutations of mechanisms represented by this 2x3 matrix (in this case, two permutations), SiMBA translated them into ordinary differential equation (ODE) models that could be optimized. Through parameter estimation, we optimize the kinetic parameters of the model, which enable us to calculate the AIC values for each model and selected the best-performing one. Table \ref{Table: Results Hypothetical} shows the optimal mechanism discovered in the first iteration, including the corresponding microkinetic model, and the AIC value which amounted to 1110.34. Figure \ref{Fig:complete_results} presents the model's fit against an arbitrary training experiment, visually illustrating its accuracy.

Following this initial step, SiMBA automatically proceeded to iteration 2, which leads to an increase of complexity by allowing an extra elementary step and an extra intermediate to be present in the modeling task. Consequently, in this iteration, the algorithm began with an empty 3x4 matrix representation of the reaction mechanism. Using the same process as in iteration 1, we identified and optimized the potential mechanisms for this iteration. To echo the point made in Section \ref{Methodology} regarding the importance of smart explorative methods, iteration 2 could generate 244,140,625 different matrix configurations; with backtracking, we only check the 31 configurations that make physical sense (~0.00001\% of all possibilities). The AIC value was again used to select the best model for iteration 2, which now amounted to an improvement to 106.28. A decrease in AIC from 1110.34 to 61.67 indicates a substantial improvement in model accuracy while maintaining parsimony.

SiMBA converged in four iterations, at which point the termination criterion was met, meaning that the informationally optimal mechanism was discovered in iteration 3. At iteration 4, no further reduction in AIC was observed, which met the termination criterion for model refinement, indicating that the added complexity did not yield better predictive accuracy. Table \ref{Table: Results Hypothetical} summarizes the best mechanism identified in each iteration, along with the corresponding microkinetic models and AIC values. Figure \ref{Fig:complete_results} illustrates the model fit for each selected mechanism against an arbitrary training experiment, further demonstrating the progressive refinement of the models across iterations (with exception of the last one).

A comparison of the final selected model against the data-generating model indicates that SiMBA successfully uncovered the underlying mechanism driving the hypothetical reaction. This case study serves as a proof-of-concept, showcasing SiMBA’s ability to generate accurate microkinetic models even in the absence of direct data on intermediates and with limited in-silico data. The results demonstrate the robustness of SiMBA in handling systems that feature both first- and second-order elementary steps, confirming its potential for more complex chemical systems and broader industrial applications, as demonstrated in the next subsections.

\subsection{The Aldol Condensation Reaction}

The aldol condensation reaction provided a different challenge for SiMBA due to the complexity of the overall system, given that there is a higher number of chemical species involved. Despite the absence of direct data on intermediates, SiMBA demonstrated its capability to infer a reliable microkinetic model, even in this constrained setting, which matches perfectly to the data-generating model.

In the first iteration, based on the stoichiometric relationship between the reactants and products, SiMBA identified a single elementary step where one mole of acetophenone (A) reacted with one mole of benzaldehyde (B) to produce chalcone (C) and water (D). The simplicity of this mechanism, represented by a 1x4 matrix (one step involving four species), aligned with the overall stoichiometry of the aldol condensation reaction and was deemed the only feasible configuration for this initial phase.

Following this, SiMBA translated the 1x4 matrix into a set of ODEs that could be optimized computationally. Parameter estimation was performed, and the AIC value was calculated and used to gauge the model's fit. Table \ref{Table: Results Aldol} does not present the mechanism identified during this initial iteration for spacing reasons, but Figure \ref{Fig:complete_results} shows how well the model predictions aligned with experimental data from a selected training experiment.

Upon completion of iteration 1, SiMBA advanced to iteration 2, where an additional elementary step and species were incorporated. This expanded the search space to a 2x5 matrix, increasing the complexity of the possible mechanisms. As in the previous step, the new sets of ODEs were optimized, and AIC values were calculated. The results from iteration 2 showed a notable improvement, as the added complexity contributed to a better overall fit, without overfitting the system. Table \ref{Table: Results Aldol} and Figure \ref{Fig:complete_results} detail the refined mechanism and its improved accuracy.

The iterative process continued, and SiMBA reached its optimal solution at iteration 3, where no further reduction in the AIC value was observed in subsequent iterations. The complexity added in iteration 4 did not yield a better AIC value, signaling that the best mechanism had already been identified, since further complexity was improving the fit negligibly. The termination criterion was therefore met after iteration 4, confirming that iteration 3 provided the most accurate and parsimonious model. The results from the second, third and fourth iterations are summarized in Table \ref{Table: Results Aldol}, while Figure \ref{Fig:complete_results} illustrates the fit for every iteration.

The comparison between the selected model and the original data-generating mechanism demonstrates SiMBA’s ability to successfully uncover the fundamental dynamics of the aldol condensation reaction, even when working with limited data. This case study serves as further evidence of SiMBA’s strength in identifying complex reaction mechanisms in realistic systems. The results not only validate SiMBA’s accuracy but also highlight its potential for broader application in mechanistic discovery across diverse chemical reactions.

\subsection{The Dehydration of Fructose}

The application of SiMBA to the dehydration of fructose case study demonstrated its ability to uncover a mechanistic pathway that aligns with literature-accepted models \cite{Chen_2024}, even though the data originated from a rate law validated by experimental findings \cite{vanPutten_2013} rather than from a constructed microkinetic model with hidden intermediates (alike the other two presented case studies). By working with a system where only the concentrations of fructose ($A$), water ($B$), and hydroxymethylfurfural ($C$) were available, SiMBA inferred the presence and behavior of unobserved species in a manner that remained consistent with a widely accepted reaction mechanism in literature.

\begin{figure}[H]
    \centering
    \includegraphics[width=1\textwidth]{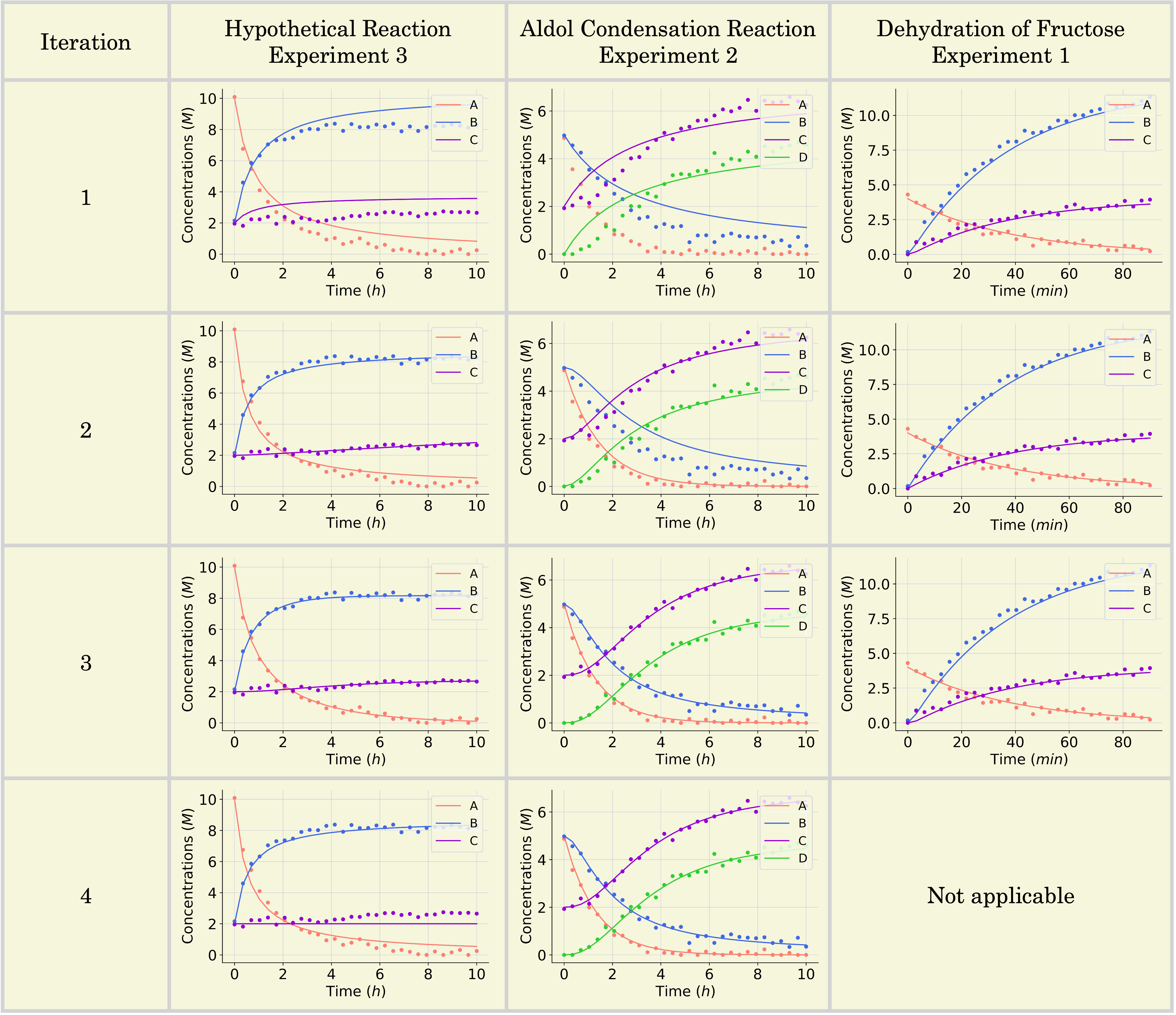}
    \caption{Model fit of the selected mechanisms across iterations for three case studies: the Hypothetical Reaction (Experiment 3), Aldol Condensation Reaction (Experiment 2), and Dehydration of Fructose Reaction (Experiment 1). Each plot shows the concentration profiles of species A (blue), B (red), C (yellow), and D (where applicable) over time, with solid lines representing model predictions and dotted points corresponding to in-silico data. For the hypothetical reaction (left column), SiMBA progressively refines the mechanism through iterations, achieving the best model fit by iteration 3. The aldol condensation reaction (middle column) shows notable improvement in fit by iteration 3, where the model captures the observed data well for all species. Iteration 4 introduces additional complexity but does not improve the fit, as demonstrated by the increased AIC value. For the dehydration of fructose reaction (right column), the model performs well from iteration 1, with iteration 2 yielding the optimal mechanism, which aligns with the data-generating model. The AIC value increases in iteration 3, signaling that additional complexity does not enhance the fit, and the process is terminated.}
    \label{Fig:complete_results}
\end{figure}

In the first iteration, SiMBA identified all permutations of the simplest possible reaction configuration consistent with the overall stoichiometry. For this case study, these permutations resulted in 10 candidate reaction matrices, each describing a minimal three-step mechanism involving five total species. The best of these initial models, shown in the first row of Table \ref{Table: Results Dehydration}, achieved an AIC value of -166.18, indicating a reasonable fit to the in-silico data. As seen in the top right plot of Figure \ref{Fig:complete_results}, this initial mechanism satisfactorily captures the concentration profiles of $A$, $B$, and $C$, yet leaves open the possibility that additional chemical complexity could yield a still better match to the observed dynamics. 

\begin{figure}[H]
    \centering
    \includegraphics[width=0.7\textwidth]{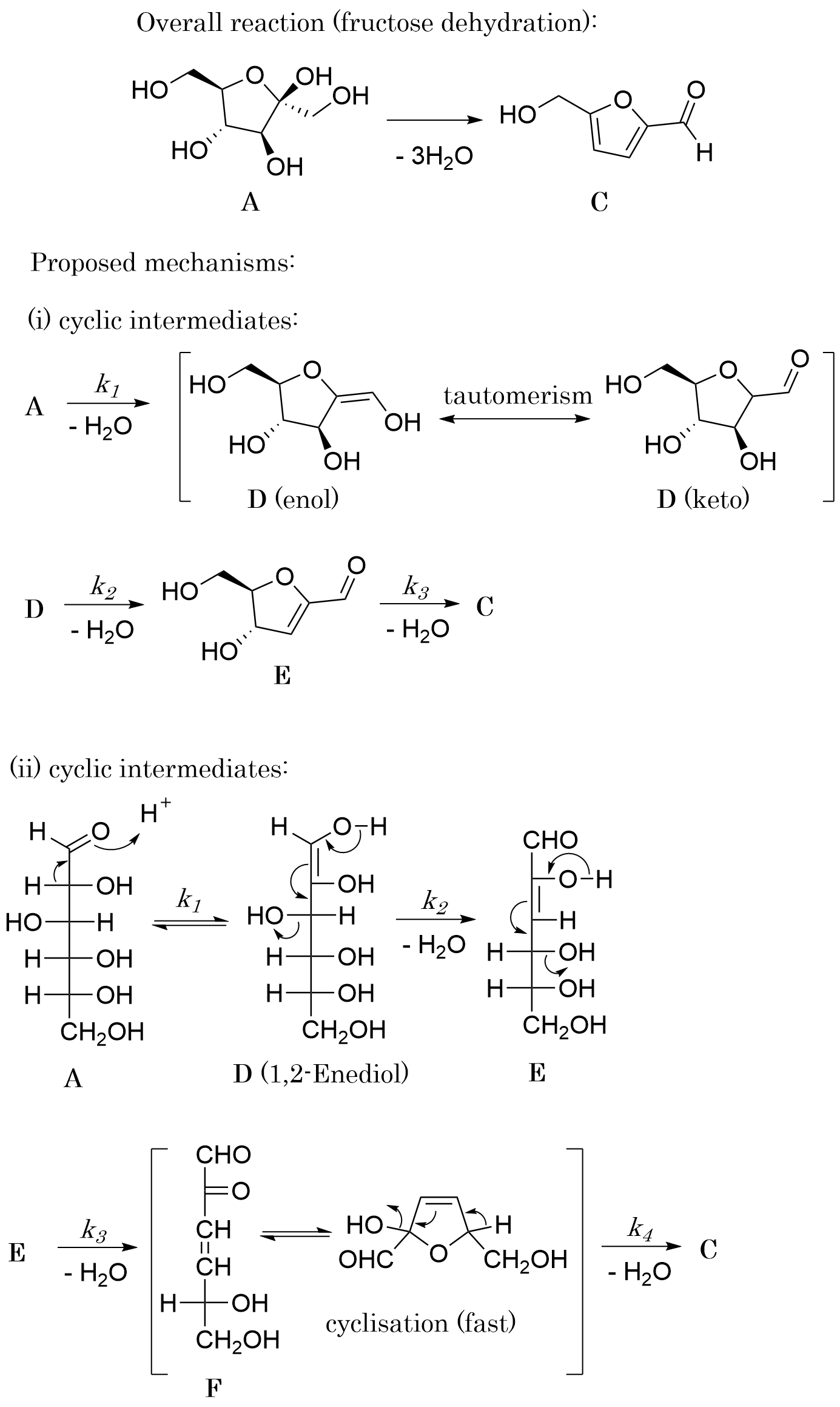}
    \caption{The transformation of fructose ($A$) to HMF ($C$) is known to be facile and involves three dehydration steps, eliminating 3 molecules of water. There are two general mechanistic pathways that are commonly proposed in literature. In the cyclic pathway (found in iteration 1), the five‐membered ring remains intact and undergoes three consecutive dehydration steps: the first step yields intermediate $D$ (enol or keto tautomer), followed by a second dehydration to produce intermediate $E$, and a final dehydration to form HMF. In the acyclic pathway (found in iteration 2 and chosen by SiMBA), fructose is proposed to adopt an open‐chain form, which tautomerizes through an enediol intermediate (also labeled $D$). After two sequential dehydration steps, the resulting intermediate $F$ cyclises readily, and the last dehydration step produces HMF. Both routes eliminate a total of three water molecules.}
    \label{Fig:fructose_chem}
\end{figure}

It is interesting to see that in iteration 1, SiMBA proposes a mechanism that is analogous to a widely accepted one in literature in which the 5-membered ring of fructose remains intact \cite{vanPutten_2013}. The first dehydration step yields intermediate $D$, which can exist either as the enol or keto tautomers. The second elimination introduces a unit of unsaturation in the ring ($E$). Finally, HMF can be obtained from the last dehydration from the ring. Given that the tautomerism is a very fast process, this mechanism can be described  kinetically by three consecutive elementary dehydration steps. The detailed mechanism can be found in Fig. \ref{Fig:fructose_chem} (i).

Iteration 2 introduced a more elaborate mechanism by appending an additional elementary step and including an extra intermediate species ($F$). The resulting 4x6 matrix improved the fit of the model, offering a more nuanced description of how fructose converts into HMF through an additional intermediate stage. As reported in Table \ref{Table: Results Dehydration}, this enhanced mechanism yielded an AIC of -167.73 -- a slight improvement that underscores a better balance between complexity and predictive accuracy. The middle-right plot of Figure \ref{Fig:complete_results} shows a slightly improved alignment between the simulated trajectories and the measured concentrations, particularly for water ($B$), which now follows the experimentally validated rate model’s curvature slightly more precisely. 

Remarkably, in this iteration, SiMBA also recovers an underlying sequence of elementary steps that is analogous to another widely accepted mechanism in the literature for the dehydration of fructose \cite{vanPutten_2013,vanPutten_2013_2}. This mechanism starts with the acyclic form of fructose ($A$), which is thought to be more abundant. In order for dehydration steps to occur, the formation of an enediol intermediate ($D$) is thought to be critical. Following two sequential dehydration steps, the dideoxyhexosulose intermediate ($F$) can cyclise very readily to form the 5-membered ring, prior to the elimination of the final water molecule to yield HMF ($C$). The detailed mechanism can be found in Fig. \ref{Fig:fructose_chem} (ii). 

Iteration 3 further expanded the proposed mechanism by introducing yet another species ($G$). Although this more complex mechanism continued to reproduce the trajectory of the reaction reasonably well, its increased complexity did not translate into further gains in predictive power; the AIC rose to -161.52, signifying that the added steps merely inflated model complexity without meaningful improvement in data fitting. This is corroborated by the bottom-right plot of Figure \ref{Fig:complete_results}, where the concentration profiles remain comparable to those from iteration 1, highlighting diminishing returns on model complexity. Consequently, the algorithm converged in iteration 3, as no additional refinement offered a better trade-off between model simplicity and accuracy. Among all iterations, the mechanism discovered in iteration 2 proved optimal. 

In this example, unlike our other case studies -- where we deliberately hid intermediates in a microkinetic simulation -- this example underscores SiMBA’s value in a realistic setting. The results establish that SiMBA can parse complex, experimentally grounded datasets and distill them into verifiable mechanistic pathways, further solidifying its robustness and practical relevance for catalytic reaction discovery.

\subsection{Methodological Limitations}\label{Limitations}
While SiMBA represents an advancement in the automated construction of microkinetic models, it is not without its limitations. One of the primary challenges is the lack of inherent chemical identification for intermediates, which requires expert input when dealing with complex systems. This limitation can constrain the algorithm's utility in scenarios where the identification of novel intermediates is crucial for understanding the reaction mechanism. Additionally, SiMBA's approach to exploring extensive mechanism spaces is computationally demanding, particularly as the complexity of the potential mechanisms increases. Furthermore, the optimization process can be sensitive to initial parameter guesses, which might lead to suboptimal solutions if not managed carefully.

To mitigate these limitations, several strategies have been implemented within the current study. For instance, to address the computational demands associated with exploring large mechanism spaces, we have employed a backtracking technique, as detailed in Section \ref{Generation}. This method significantly reduces the search space by eliminating unfeasible pathways early in the process, and the exploration has been parallelized to further improve computational efficiency. To counter the potential sensitivity to initial parameter guesses during optimization, as discussed in Section \ref{Estimation}, we have utilized a well-established optimization algorithm, specifically the BFGS algorithm, with a multi-start option. This approach increases the likelihood of finding the global optimum by starting the optimization from multiple initial guesses.

Looking ahead, future work will focus on overcoming the lack of inherent chemical knowledge as well as continuing to reduce the computational cost to further enhance SiMBA's capabilities. For the issue of chemical identification of intermediates, we plan to explore the use of large language models (LLMs) and their auto-completion capabilities for mechanisms. These models have shown success in completing partial reaction networks \cite{vanWijngaarden_2024}, and integrating them into SiMBA could provide a more autonomous solution for intermediate identification, reducing the need for expert intervention. Additionally, we are considering the integration of uncertainty quantification methods, which will increase the robustness of the models proposed by SiMBA. These enhancements aim to make SiMBA a powerful tool for chemists and engineers, capable of addressing the diverse challenges encountered in kinetic discovery and reaction mechanism elucidation.

\section{Conclusions}\label{Conclusions}
In this paper, we have presented SiMBA (Simplest Mechanism Builder Algorithm), an efficient approach to microkinetic model discovery that aims to address key limitations in both manual and automated methods. Microkinetic models play a crucial role in various industries, including pharmaceuticals, petrochemicals, and environmental engineering, by helping to optimize chemical processes and understanding their environmental impact. However, traditional methods for constructing these models are often time-consuming, complex, and prone to human error, as they require extensive expertise to manually identify reaction mechanisms and intermediates. While automated approaches have emerged to overcome these challenges, they tend to generate overly complex models or rely heavily on prior knowledge, limiting their practical application.

SiMBA was developed to fill this gap by introducing a minimalistic, data-driven approach that incrementally builds model complexity based on available information. Unlike other methods, SiMBA begins with the simplest possible mechanism and systematically adds complexity only if the additional parameters provide informational gain. This balance between simplicity and accuracy is achieved through four key phases: mechanism generation, mechanism translation, parameter estimation, and model comparison. The algorithm starts by proposing feasible reaction mechanisms using a parallelized backtracking algorithm, translates these mechanisms into systems of ODEs, optimizes their kinetic parameters, and selects the best model using the AIC to ensure the right trade-off between model complexity and fit.

The effectiveness of SiMBA was demonstrated through three case studies: a hypothetical reaction, an aldol condensation, and the dehydration of fructose. In each case, SiMBA successfully distilled complex reaction behaviors into accurate models, even in situations where intermediates were not directly observable. These case studies highlight the algorithm’s versatility and robustness in generating models.

While SiMBA has proven to be a powerful tool for microkinetic model discovery, it is not without limitations. The current version does not provide chemical identification of intermediates, necessitating expert input for it. Additionally, while the algorithm excels at balancing simplicity with accuracy, incorporating uncertainty quantification could further enhance the robustness of its predictions. Future work will focus on integrating more chemical knowledge and techniques for identifying intermediates, as well as expanding the algorithm's capabilities to address uncertainty in model predictions.

In conclusion, SiMBA offers a novel approach to overcoming many of the challenges associated with existing automated methods for microkinetic model discovery. By systematically generating, refining, and evaluating microkinetic models, SiMBA provides a new framework for mechanistic discovery. As SiMBA continues to evolve with future enhancements like uncertainty quantification and intermediate identification, we hope that it will become a useful tool for chemists and engineers, helping bridge the gap between theoretical exploration and industrial applications. 

\begin{table}[H]
\caption{Reaction mechanisms, corresponding microkinetic models, and AIC values for iterations 1 through 4. The table presents the reaction mechanisms discovered at each iteration, alongside their respective microkinetic models and AIC values. Each iteration reflects an increase in the complexity of the reaction mechanism. In iteration 1, the simplest model consists of two elementary steps involving species A, B, and C, with an AIC value of 1139.86. By iteration 2, an additional intermediate (D) is introduced, lowering the AIC to 106.28. The optimal mechanism (in this case, identical to the data-generating one), discovered in iteration 3, involves the introduction of an additional intermediate (E), achieving the lowest AIC of -317.99. Iteration 4 introduces yet another intermediate (F), but results in a higher AIC value of 390.92, indicating that iteration 3 provides the best balance between accuracy and complexity.\\}
\begin{tabular}{llll}
\hline
    Iteration & Reaction Mechanism & Microkinetic Model & AIC Value \\
\hline
    1 & $\begin{matrix}
    2A \xrightarrow{k_1} B \\
    2A \xrightarrow{k_2} C
    \end{matrix}$ & $\begin{aligned}
    \\
    &\frac{dC_A}{dt} = -k_1C_A^2 - k_2C_A^2 \\
    &\frac{dC_B}{dt} = k_1C_A^2 \\
    &\frac{dC_C}{dt} = k_2C_A^2 \\
    \\
    \end{aligned}$ & 1139.86 \\
    \hline
    2 & $\begin{matrix}
    2A \xrightarrow{k_1} D \\
    2A \xrightarrow{k_2} B \\
    D \xrightarrow{k_3} C
    \end{matrix}$ & $\begin{aligned}
    \\
    &\frac{dC_A}{dt} = -k_1C_A^2 - k_2C_A^2 \\
    &\frac{dC_B}{dt} = k_2C_A^2 \\
    &\frac{dC_C}{dt} = k_3C_D \\
    &\frac{dC_D}{dt} = -k_3C_D \\
    \\
    \end{aligned}$ & 106.28 \\
    \hline
    3 & $\begin{matrix}
    2A \xrightarrow{k_1} B \\
    A \xrightarrow{k_2} D \\
    D \xrightarrow{k_3} E \\
    A + E \xrightarrow{k_4} C
    \end{matrix}$ & $\begin{aligned}
    \\
    &\frac{dC_A}{dt} = -k_1C_A^2 - k_2C_A - k_4C_AC_E \\
    &\frac{dC_B}{dt} = k_1C_A^2 \\
    &\frac{dC_C}{dt} = k_4C_AC_E \\
    &\frac{dC_D}{dt} = k_2C_A + k_3C_D \\
    &\frac{dC_E}{dt} = k_3C_D - k_4C_AC_E \\
    \\
    \end{aligned}$ & -317.99 \\
    \hline
    4 & $\begin{matrix}
    2A \xrightarrow{k_1} B \\
    2A \xrightarrow{k_2} D \\
    D \xrightarrow{k_3} 2E \\
    E \xrightarrow{k_4} F \\
    E + F \xrightarrow{k_5} C
    \end{matrix}$ & $\begin{aligned}
    \\
    &\frac{dC_A}{dt} = -k_1C_A^2 - k_2C_A^2 \\
    &\frac{dC_B}{dt} = k_1C_A^2 \\
    &\frac{dC_C}{dt} = k_5C_EC_F \\
    &\frac{dC_D}{dt} = k_2C_A^2 - k_3C_D \\
    &\frac{dC_E}{dt} = k_3C_D - k_4C_E - k_5C_EC_F \\ 
    &\frac{dC_F}{dt} = k_4C_E - k_5C_EC_F \\
    \\
    \end{aligned}$ & 390.92 \\
\hline
\end{tabular}
\label{Table: Results Hypothetical}
    \centering
\end{table}

\begin{table}[H]
\caption{Reaction mechanisms, microkinetic models, and AIC values for iterations 2 to 4 identified by SiMBA in the aldol condensation case study. Due to space constraints, iteration 1 is omitted from the table but its performance is shown in Figure \ref{Fig:complete_results}. The table presents the reaction mechanisms for iterations 2 through 4, the corresponding microkinetic models in the form of ordinary differential equations, and the AIC values, which assess model quality. Iteration 2 begins with the introduction of intermediate E, resulting in a substantial improvement in the AIC value to 866.58. By iteration 3, the inclusion of intermediate F yields the optimal model with the lowest AIC value of -351.17, indicating the best balance between fit and parsimony. In iteration 4, an additional intermediate (G) is introduced, but the AIC value of -349.21 indicates that the added complexity is unnecessary, suggesting that the optimal model was discovered in iteration 3.}
\begin{tabular}{llll}
\hline
    Iteration & Reaction Mechanism & Microkinetic Model & AIC Value \\
\hline
    2 & $\begin{matrix}
    A \xrightarrow{k_1} E \\
    B + E \xrightarrow{k_2} C + D
    \end{matrix}$ & $\begin{aligned}
    \\
    &\frac{dC_A}{dt} = -k_1C_A \\
    &\frac{dC_B}{dt} = -k_2C_BC_E \\
    &\frac{dC_C}{dt} = k_2C_BC_E \\
    &\frac{dC_D}{dt} = k_2C_BC_E \\
    &\frac{dC_E}{dt} = k_1C_A - k_2C_BC_E \\
    \\
    \end{aligned}$ & 866.58 \\
    \hline
    3 & $\begin{matrix}
    A \xrightarrow{k_1} E \\
    B + E \xrightarrow{k_2} F \\
    F \xrightarrow{k_3} C + D
    \end{matrix}$ & $\begin{aligned}
    \\
    &\frac{dC_A}{dt} = -k_1C_A \\
    &\frac{dC_B}{dt} = -k_2C_BC_E \\
    &\frac{dC_C}{dt} = k_3C_F \\
    &\frac{dC_D}{dt} = k_3C_F \\
    &\frac{dC_E}{dt} = k_1C_A - k_2C_BC_E \\
    &\frac{dC_F}{dt} = k_2C_BC_E - k_3C_F \\
    \\
    \end{aligned}$ & -351.17 \\
    \hline
    4 & $\begin{matrix}
    A \xrightarrow{k_1} E \\
    E + B \xrightarrow{k_2} F + G \\
    F \xrightarrow{k_3} C \\
    G \xrightarrow{k_4} D
    \end{matrix}$ & $\begin{aligned}
    \\
    &\frac{dC_A}{dt} = -k_1C_A\\
    &\frac{dC_B}{dt} = -k_2C_BC_E \\
    &\frac{dC_C}{dt} = k_3C_F \\
    &\frac{dC_D}{dt} = k_4C_G \\
    &\frac{dC_E}{dt} = k_1C_A - k_2C_BC_E \\ 
    &\frac{dC_F}{dt} = k_2C_BC_E - k_3C_F \\
    &\frac{dC_G}{dt} = k_2C_BC_E - k_4C_G \\
    \\
    \end{aligned}$ & -349.21 \\
\hline
\end{tabular}
\label{Table: Results Aldol}
    \centering
\end{table}

\begin{table}[H]
\caption{Evolution of reaction mechanisms, microkinetic models, and AIC values across three iterations of the SiMBA process for the dehydration of fructose case study. The table shows the progression of SiMBA through iterations 1, 2, and 3 for the given reaction system. For each iteration, the reaction mechanism, corresponding microkinetic model, and the AIC values are presented. In iteration 1, the simplest mechanism is identified with an AIC of -166.18. As complexity increases in iterations 2 and 3, intermediates such as F and G are introduced, and the model structure becomes more intricate. The best-fit mechanism is achieved in iteration 2 with an AIC of -167.73, while iteration 3, despite introducing additional complexity, yields a higher AIC value of -161.52, indicating that further refinement may not improve model accuracy considerably.}
\begin{tabular}{llll}
\hline
    Iteration & Reaction Mechanism & Microkinetic Model & AIC Value \\
\hline
    1 & $\begin{matrix}
    A \xrightarrow{k_1} B + D \\
    D \xrightarrow{k_2} B + E \\
    E \xrightarrow{k_3} B + C
    \end{matrix}$ & $\begin{aligned}
    \\
    &\frac{dC_A}{dt} = -k_1C_A \\
    &\frac{dC_B}{dt} = k_1C_A + k_2C_D + k_3C_E \\
    &\frac{dC_C}{dt} = k_3C_E \\
    &\frac{dC_D}{dt} = k_1C_A - k_2C_D \\
    &\frac{dC_E}{dt} = k_2C_D - k_3C_E \\
    \\
    \end{aligned}$ & -166.18 \\
    \hline
    2 & $\begin{matrix}
    A \xrightarrow{k_1} D \\
    D \xrightarrow{k_2} B + E \\
    E \xrightarrow{k_3} B + F \\
    F \xrightarrow{k_4} B + C
    \end{matrix}$ & $\begin{aligned}
    \\
    &\frac{dC_A}{dt} = -k_1C_A \\
    &\frac{dC_B}{dt} = k_2C_D + k_3C_E + k_4C_F \\
    &\frac{dC_C}{dt} = k_4C_F \\
    &\frac{dC_D}{dt} = k_1C_A - k_2C_D \\
    &\frac{dC_E}{dt} = k_2C_D - k_3C_E \\
    &\frac{dC_F}{dt} = k_3C_E - k_4C_F \\
    \\
    \end{aligned}$ & -169.40 \\
    \hline
    3 & $\begin{matrix}
    A \xrightarrow{k_1} B + D \\
    D \xrightarrow{k_2} E + F \\
    E + F \xrightarrow{k_3} D \\
    D \xrightarrow{k_4} B + G \\
    G \xrightarrow{k_5} B + C
    \end{matrix}$ & $\begin{aligned}
    \\
    &\frac{dC_A}{dt} = -k_1C_A \\
    &\frac{dC_B}{dt} = k_1C_A + k_4C_D + k_5C_G \\
    &\frac{dC_C}{dt} = k_5C_G \\
    &\frac{dC_D}{dt} = k_1C_A - k_2C_D + k_3C_EC_F - k_4C_D \\
    &\frac{dC_E}{dt} = k_2C_D - k_3C_EC_F \\
    &\frac{dC_F}{dt} = k_2C_D - k_3C_EC_F \\
    &\frac{dC_G}{dt} = k_4C_D - k_5C_G \\
    \\
    \end{aligned}$ & -161.52\\
\hline
\end{tabular}
\label{Table: Results Dehydration}
    \centering
\end{table}

\section*{Author Contributions}

\textbf{Miguel Ángel de Carvalho Servia:} Conceptualization, formal analysis, investigation, methodology, project administration, software development, validation, visualization, writing (original draft), and writing (review and editing).

\textbf{King Kuok (Mimi) Hii:} Conceptualization, formal analysis, funding acquisition, supervision, writing (original draft), and writing (review and editing).

\textbf{Klaus Hellgardt:} Conceptualization, formal analysis, funding acquisition, supervision, and writing (review and editing).

\textbf{Dongda Zhang:} Conceptualization, funding acquisition and supervision.

\textbf{Ehecatl Antonio del Rio Chanona:} Conceptualization, formal analysis, funding acquisition, methodology, project administration, supervision, and writing (review and editing).

\section*{Declaration of Competing Interest}
The authors declare that they have no known competing financial interests or personal relationships that could have appeared to influence the work reported in this paper.

\section*{Acknowledgments and Funding}
This work was supported by the EPSRC Centre of Doctoral Training for Next Generation Synthesis \& Reaction Technology (rEaCt) funding grant EP/S023232/1.

\section*{Appendix A. Supplementary Information}


The code used to produce all results and graphs shown in this work can be accessed at \url{https://github.com/OptiMaL-PSE-Lab/auto_react_mech_construct}.




\bibliographystyle{unsrtnat}
\bibliography{references.bib}

\end{document}